\documentclass[12pt]{iopart}
\usepackage{amsfonts}
\newcommand{\ex}{\frac{\rme^{\rmi(t'-t)(E_1-E_2)/\la^2}}{\la^2}}
\newcommand{\ga}{\gamma}
\newcommand{\la}{\lambda}
\newcommand{\om}{\otimes}
\newcommand{\ep}{\varepsilon}
\newcommand{\dl}{\delta}
\newcommand{\ba}{\begin{array}{l}}
\newcommand{\ea}{\end{array}}
\newtheorem{theorem}{Theorem}

\begin{document}

\title[Quantum stochastic equation]
      {Quantum stochastic equation for the low density limit}

\author{L Accardi\dag, A N Pechen\dag\footnote[3]{permanent address:
       Steklov Mathematical Institute of Russian Academy of
       Sciences, Gubkin St.8, GSP-1, 117966, Moscow, Russia}
       and I V Volovich\ddag}

\address{\dag\ Centro Vito Volterra, Universita di Roma Tor Vergata 00133,Italia}

\address{\ddag\ Steklov Mathematical Institute of Russian Academy of
        Sciences, Gubkin St.8, GSP-1, 117966, Moscow, Russia}

\eads{\mailto{accardi@volterra.mat.uniroma2.it}, \mailto{pechen@mi.ras.ru},
      \mailto{volovich@mi.ras.ru}}

\begin{abstract}
A new derivation of quantum stochastic differential equation
for the evolution operator in the low density limit is presented.
We use the distribution approach and derive a new algebra for
quadratic master fields in the low density limit by using
the energy representation. We formulate the stochastic golden rule
in the low density limit case
for a system coupling with Bose field via quadratic interaction.
In particular the vacuum expectation value of the evolution
operator is computed and its exponential decay is shown.

\end{abstract}

\submitto{Journal of Physics A: Mathematical and General}


\maketitle

%%%%%%%%%%%%%%%%%%%%%%%%%%%%%%%%%%%%%%%%%%%%%%%%%%%%%%%%%%%%%%%%%%%%%%%%%

\section{Introduction}

There are many studies of the large time behavior in quantum
theory. One of the powerful methods is the stochastic limit. Many
important physical models have been investigated by using this
method (see \cite{AFL,AKV,ALV} for more discussions). This method,
however, is restricted to the studying of the long time behavior
of the models in the weak coupling case, i.e. when coupling
constant is a small parameter, and it can not be applied directly
to important class of models which contain terms in the
interaction without small coupling constant.

This last class of
models includes models in which the small parameter is the
density. Such models naturally arise in the low density limit
(LDL). The LDL for a classical Lorentz gas is the Boltzmann-Grad limit.
For classical systems there has been a considerable progress in
the rigorous derivation of the Boltzmann equation.
Lanford~\cite{lanford} using ideas of Grad~\cite{grad} proved the
convergence of the hierarchy of correlation functions for a hard
sphere gas in the Boltzmann--Grad limit for sufficiently short
times. This proof was extended by King~\cite{king} to positive
potentials of finite range. The limiting evolution of the one
particle distribution is governed by the non-linear Boltzmann
equation.

The test particle problem was studied by many authors(see
a review of Spohn~\cite{spohn2}). One considers
the motion of a single particle through an environment of randomly
placed, infinitely heavy scatterers (Lorentz gas). In the
Boltzmann--Grad limit successive collisions become independent and
the averaged over the positions of the scatterers the position and
velocity distribution of the particle converges to the solution of
the linear Boltzmann equation.

To describe a quantum physical model to which LDL can be applied
let us consider an N-level atom immersed in a free gas whose
molecules can collide with the atom; the gas is supposed to be
very dilute. Then the reduced time evolution for the atom will be
Markovian, since the characteristic time  $t_S$  for appreciable
action of the surroundings on the atom (time between collisions)
is much larger than the characteristic time  $t_R$  for relaxation
of correlations in the surroundings. Rigorous results
substantiating this idea have been obtained in~\cite{dumcke}.

It is known~\cite{AcLu} that the dynamics of the N-level atom
interacting with the free gas converges, in the low density limit,
to the solution of a quantum stochastic differential equation
driven by quantum Poisson noise. Indeed, from a semiclassical
point of view, collision times, being times of occurrence of rare
events, will tend to become Poisson distributed, whereas the
effect of each collision will be described by the
(quantum-mechanical) scattering operator of the atom with one gas
particle (see the description of the quantum Poisson process
in~\cite{kumerrer}).

In this paper a derivation of the quantum stochastic differential
equation for the evolution operator in the low density limit is
presented. The equation obtained is equivalent to the stochastic
equation which has been derived in \cite{AcLu} but we use a new
method. We use the distribution approach~\cite{AKV,ALV}
and derive a new algebra for quadratic master
fields in the low density limit by using the energy
representation. An advantage of this method is the simplicity of
derivation of quantum stochastic equations and computation of
correlation functions. We formulate the stochastic golden rule in
the low density limit case for a system coupling with Bose field
via quadratic interaction. In particular the vacuum expectation
value of the evolution operator is computed and its exponential
decay is shown.

Main results of the paper are quantum stochastic differential
equation~(\ref{normordeq}),
the new algebra of commutation relations for the master
field (Theorem~\ref{mastfi}) and the expression for the expectation value
of the evolution operator~(\ref{expvalue}).

In this paper we obtain unitary evolution which is given by
the solution of the quantum stochastic differential equation.
Using this equation one can obtain corresponding quantum
Langevin and master equations.

An important problem in theory of open quantum systems is the
rigorous derivation of quantum Boltzmann equation from microscopic
dynamics. Aa approach to the derivation of classical and quantum
Boltzmann equations, based on BBGKY-hierarchy, has been presented
in the work of Bogoliubov~\cite{bogol}. The low density limit for the model
under consideration, with completely different
methods, based on quantum BBGKY hierarchy has been investigated by
D\"umcke~\cite{dumcke}.

Let us explain our notations. We consider a quantum model of
a test particle interacting with a reservoir (heat bath). We
shall restrict ourselves to the case of a Boson reservoir in this
paper. Let $ {\cal H}_{\rm S} $ be the Hilbert space of the system
(test particle) with the system Hamiltonian $ H_{\rm
S} $. The system Hilbert space for the $N$-level atom ${\cal
H}_{\rm S} = \mathbb C^N$. The reservoir is described by the
Boson Fock space $\Gamma({\cal H}_1)$ over the one particle
Hilbert space ${\cal H}_1 = L^2(\mathbb R^d)$, where $d=3$ in
physical case. Moreover, the Hamiltonian of the reservoir is
given by $H_{\rm R}:=\rmd\Gamma(H_1)$ (the second quantization of
the one particle Hamiltonian $ H_1 $) and the total Hamiltonian
of the compound system is given by a self--adjoint operator
on the total Hilbert space ${\cal H}_{\rm S}\otimes \Gamma({\cal
H}_1)$, which has the form
\[
 H_{\rm tot}:=H_{\rm S}\otimes1+1\otimes H_{\rm R}+H_{\rm int}=:H_{\rm free} +
 H_{\rm int}.
\]
Here $H_{\rm int}$  is the interaction Hamiltonian between the
system and the reservoir.
The evolution operator at time $t$ is given by:
\[
 U_t:=\rme^{\rmi tH_{\rm free}}\cdot \rme^{-\rmi tH_{\rm tot}}.
\]
Obviously, it satisfies the differential equation
\[
 \partial_tU_t=-\rmi H_{\rm int}(t)U_t
\]
where the quantity $H_{\rm int}(t)$ will be
called the evolved interaction and defined as
\[
 H_{\rm int}(t)=\rme^{\rmi tH_{\rm free}} H_{\rm int}\rme^{-\rmi tH_{\rm free}}.
\]

The interaction Hamiltonian will be assumed to have the following
form:
\[
 H_{\rm int}:= D\otimes A^+(g_0)A(g_1)+D^+\otimes A^+(g_1)A(g_0)
\]
where $ D $ is a bounded operator in $ {\cal H}_S $, $ D\in {\bf
B}({\cal H}_S) $, $ A $ and $ A^+ $ are annihilation and creation
operators and $ g_0, g_1\in {\cal H}_1$ are form-factors
describing the interaction of the system with the reservoir.
This Hamiltonian preserves the particle number of the reservoir,
and therefore the particles of the reservoir are only scattered
and not created or destroyed. This model was considered by
Davies~\cite{davies} in the analysis of the weak coupling limit.
The development of the method to the Bose gas is a subject
of further works.

With the notion
\[
 S^0_t:=\rme^{\rmi tH_1} \ ;\qquad D(t):=\rme^{\rmi tH_{\rm S}}D\rme^{-\rmi tH_{\rm S}}
\]
the evolved interaction can be written in the form
\begin{equation}\fl
 H_{\rm int}(t):=D(t)\otimes A^+(S^0_tg_0)A(S^0_tg_1)
 +D^+(t)\otimes A^+(S^0_tg_1)A(S^0_tg_0)\label{Hint}
\end{equation}

The initial state of the compound system is supposed to be of
the form
\[
 \rho = \rho_{\rm S}\otimes\varphi^{(\xi)}.
\]
Here $\rho_{\rm S}$ is arbitrary density matrix of the system and
the initial state of the reservoir is the Gibbs state, at inverse
temperature $\beta$, of the free evolution, i.e. the gauge
invariant quasi-free state $\varphi^{(\xi)}$, characterized by
\begin{equation}\label{state}
 \varphi^{(\xi)}(W(f))=\exp\Bigl(-{1\over 2}<f,(1+\xi\rme^{-\beta H_1})
 (1-\xi\rme^{-\beta H_1})^{-1}f>\Bigr)
\end{equation}
for each $f\in {\cal H}_1$. Here $ W(f) $ is the Weyl operator,
$\beta$ the inverse temperature of the reservoir,
$\xi=\rme^{\beta\mu}$ the fugacity, $\mu$ the chemical potential.
We suppose that the temperature $\beta^{-1}>0$. Therefore for
sufficiently low density one is above the transition temperature,
and no condensate is present. The generalization to the case then
the condensate is present, is a subject of further investigations.

We will study the dynamics, generated by the
Hamiltonian~(\ref{Hint}) and the initial state of the
reservoir~(\ref{state}) in the low density regime: $n\to 0$,
$t\sim 1/n$ ($n$ is the density of particles of the reservoir). In
the low density limit the fugacity $\xi$ and the density of
particles of the reservoir $n$ have the same asymptotic, i.e.
\[
 \lim\limits_{n\to 0}\frac{\xi(n)}{n}=1
\]
Therefore the limit $n\to 0$ is equivalent to the limit $\xi\to
0$.

The low density limit for this model, with completely different
methods, based on quantum BBGKY hierarchy has been considered by
D\"umcke~\cite{dumcke}.

Throughout the paper, for simplicity, the following  technical
condition is assumed: the two test functions in the interaction
Hamiltonian have disjoint supports in the energy representation.
Thus the disjointness is invariant under the action of any
function of $H$. This assumption means that the two test function
$g_0, g_1$ in the interaction Hamiltonian satisfy:
\[
 <g_0,S^0_t\rme^{-\beta H}g_1>=0\qquad \forall t\in{\mathbb R}.
\]
This means that, even if the particles of the reservoir have
generically a continuous energy spectrum, they behave like a
2--level system as far as their interaction with the system is
concerned: if $P_0$ and $P_1$ project onto disjoint intervals
(energy bands) $I_0$ and $I_1$, these energy bands act as the
counterpart of the energy levels $\epsilon_0,\epsilon_1$ of the
system.

Also the rotating wave approximation condition will be assumed.
This condition means that
\[
 \rme^{\rmi tH_{\rm S}}D\rme^{-\rmi tH_{\rm S}}=\rme^{-\rmi t\omega_0}D
\]
where $ \omega_0 $ is a real number. This is a familiar
assumption, satisfied by all the Hamiltonians commonly used in
quantum optics. This assumption is satisfied if
$D=|\epsilon_0><\epsilon_1|$, where $|\epsilon_0>$ and $
|\epsilon_1>$ are eigenvectors of the free system Hamiltonian with
eigenvalues $\epsilon_0$ and $\epsilon_1$ so that $\omega =
\epsilon_1-\epsilon_0$.

We will fix a projection operator $ P_0 $ in $ {\cal H}_1 $
commuting with $ H_1 $ and $ H $ and such that
\[
 P_0g_0=g_0\quad {\rm and}\quad P_0g_1=0.
\]
Using this projection let us define the group $ \{S_t;\, t\in {\mathbb R}\} $
of unitary operators on $ {\cal H}_1 $ by
\[
 S_t=S^0_t\rme^{-\rmi t\omega_0P_0}=\rme^{\rmi t(H_1-\omega_0P_0)}.
\]
The infinitesimal generator $ H_1' $ of $ S_t $ is given by
\[
 H_1'=H_1-\omega_0P_0.
\]

Following Palmer~\cite{Pl} we realize the representation space
as the tensor product of a Fock and anti-Fock representations.
Then the expectation values with respect to the state
$ \varphi^{(\xi)} $ for the
model with the interaction Hamiltonian~(\ref{Hint}) can be
conveniently represented as the vacuum expectation values in the
Fock-anti-Fock representation for the modified Hamiltonian.

Denote by ${\cal H}_1^\iota$ the conjugate of ${\cal H}_1$, i.e.
\begin{eqnarray}
 \iota :\ {\cal H}_1\longrightarrow {\cal H}_1\ ,\ \ \iota
 (\la f):=\bar \la \iota (f) \nonumber \\
 <\iota(f),\iota(g)>_\iota \ :=\ <g,f> \nonumber
\end{eqnarray}
then, ${\cal H}_1^\iota $ is a Hilbert space.
The corresponding Fock space $ \Gamma({\cal H}_1^\iota) $ is called
in this context the anti-Fock space.

It was shown in~\cite{AcLu}
that with notations $ D_0=D,\, D_1=D^+ $ the modified Hamiltonian acting in
$ \Gamma({\cal H}_1)\otimes \Gamma({\cal H}_1^\iota) $ has the form
\begin{eqnarray*}\fl
 H_\la(t)=\sum\limits_{\ep=0,1}D_\ep\otimes\bigl(A^+(S_tg_\ep)A(S_tg_{1-\ep})
 \otimes 1\\
 +\la(A(S_tg_{1-\ep})\otimes
 A(S_tLg_\ep)+A^+(S_tg_\ep)\otimes A^+(S_tLg_{1-\ep})\bigr).
\end{eqnarray*}
Here $ A $ and $ A^+ $ are Bose annihilation and creation operators
acting in the Fock spaces $ \Gamma({\cal H}_1) $ and
$ \Gamma({\cal H}_1^\iota) $ and $ L:=\rme^{-\beta H/2} $.

Moreover, it will be assumed that there exists a subset ${\cal K}$
(which includes $g_0, g_1$) of the one particle Hilbert space ${\cal
H}_1$, such that
\[
 \int_{\mathbb R}|<f,S_tg>|dt<\infty\qquad \forall f,g\in {\cal K}.
\]

The interaction Hamiltonian determines the evolution operator $ U_t^{(\la)} $
which is the solution of the Schr\"odinger equation
in interaction representation:
\[
\partial_tU_t^{(\la)}=-\rmi H_\la(t)U_t^{(\la)}
\]
with initial condition
\[
U_0^{(\la)}=1.
\]
One has the following integral equation for the evolution operator.
\[
U_t^{(\la)}=1-\rmi\int\limits_0^t\rmd t'H_\la(t')U_{t'}^{(\la)}.
\]

%%%%%%%%%%%%%%%%%%%%%%%%%%%%%%%%%%%%%%%%%%%%%%%%%%%%%%%%%%%%%%%%%%%%%%%

\section{Energy representation}

We will investigate the limit of the evolution operator when $ \xi\to +0 $
after the time rescaling $ t\to t/\xi $, where $ \xi=\la^2 $.
After this time rescaling
the equation for the evolution operator becomes
\[
 \partial_tU_{t/\la^2}^{(\la)}=-\rmi\sum\limits_{\ep=0,1}D_\ep\om
 \bigl(N_{\ep,1-\ep,\la}(t)+B_{1-\ep,\ep,\la}(t)+B^+_{\ep,1-\ep,\la}(t)\bigr)
 U^{(\la)}_{t/\la^2}
\]
where we introduced the notations:
\begin{eqnarray*}
 N_{\ep_1,\ep_2,\la}(t)=\frac{1}{\la^2}A^+(S_{t/\la^2}g_{\ep_1})
 A(S_{t/\la^2}g_{\ep_2})\om 1 \\
 B^+_{\ep_1,\ep_2,\la}(t)=\frac{1}{\la}
 A^+(S_{t/\la^2}g_{\ep_1})\om A^+(S_{t/\la^2}Lg_{\ep_2}).
\end{eqnarray*}

Let us introduce the energy representation for the creation
and annihilation operators by the formulae
\begin{equation}\label{AEdef}\fl
 A^+_{\rm E}(g)=A^+(P_{\rm E}g)=\int\rmd k(P_Eg)(k)a^+(k)=\int\rmd k\dl(H_1'-E)g(k)a^+(k)
\end{equation}
\[
 A_E(g)=A(P_Eg)
\]
Here
\[
 P_E=\frac{1}{2\pi}\int\limits_{-\infty}^{\infty}\rmd tS_t\rme^{-\rmi tE}=
 \dl(H_1'-E).
\]
It has the properties
\begin{eqnarray*}
 S_t=\int\rmd EP_E\rme^{\rmi tE} \\
 P_EP_{E'}=\dl(E-E')P_E \\
 P^*_E=P_E
\end{eqnarray*}

In the case when $ {\cal H}_1=L^2({\mathbb R}^d) $ the one-particle
Hamiltonian is the multiplication operator to the function
$ \omega(k) $ and acts on any element $ f\in L^2({\mathbb R}^d) $ as
$ H_1f(k)=\omega(k)f(k) $.

It is easy to check that
\[
 [A_E(f),A_{E'}^+(g)]=\dl(E-E')<f,P_Eg>.
\]
Here $ <\cdot ,\cdot > $ means the scalar product in $ {\cal H}_1 $.

Using the energy representation one gets
\begin{eqnarray*}
 N_{\ep_1,\ep_2,\la}(t)=\int\rmd E_1\rmd E_2N_{\ep_1,\ep_2,\la}(E_1,E_2,t) \\
 B_{\ep_1,\ep_2,\la}(t)=\int\rmd E_1\rmd E_2B_{\ep_1,\ep_2,\la}(E_1,E_2,t)
\end{eqnarray*}
where
\begin{eqnarray*}
 N_{\ep_1,\ep_2,\la}(E_1,E_2,t):=\frac{\rme^{\rmi t(E_1-E_2)/\la^2}}{\la^2}
 A^+_{E_1}(g_{\ep_1})A_{E_2}(g_{\ep_2})\om 1 \\
 B_{\ep_1,\ep_2,\la}(E_1,E_2,t):=
 \frac{\rme^{\rmi t(E_2-E_1)/\la^2}}{\la}A_{E_1}(g_{\ep_1})\om A_{E_2}(Lg_{\ep_2}).
\end{eqnarray*}

Let us also denote
\[
 \ga_\ep(E):=\int\limits^0_{-\infty}\rmd t<g_\ep,S_tg_\ep>\rme^{-\rmi tE}.
\]

%%%%%%%%%%%%%%%%%%%%%%%%%%%%%%%%%%%%%%%%%%%%%%%%%%%%%%%%%%%%%%%%%%%%%%%%%%%%%%%

\section{The limiting commutation relations}

Besides the operators $ B_{\ep_1,\ep_2,\la}(E_1,E_2,t) $ and $
N_{\ep_1,\ep_2,\la}(E_1,E_2,t) $ defined above let us consider the
following operators:
\[
 N^{\iota}_{\ep_1,\ep_2,\la}(E_1,E_2,t)=
 \frac{\rme^{\rmi t(E_2-E_1)/\la^2}}{\la^2}
 \phantom{i}1\om A^+_{E_1}(Lg_{\ep_1})A_{E_2}(Lg_{\ep_2})
\]
with $ A^+_E(Lg_\ep) $ has been defined in (\ref{AEdef}).
The commutators of these operators are:

\begin{eqnarray}
 \fl [B_{\ep_1,\ep_2,\la}(E_1,E_2,t),B^+_{\ep_3,\ep_4,\la}(E_3,E_4,t')]\nonumber\\
 \fl = \ex\Bigl(\dl_{\ep_1,\ep_3}\dl_{\ep_2,\ep_4}\dl(E_1-E_3)\dl(E_2-E_4)
 <g_{\ep_1},P_{E_1}g_{\ep_1}><g_{\ep_2},P_{E_2}L^2g_{\ep_2}>\nonumber \\
 \vphantom{\ex}
 +\la^2\dl_{\ep_1,\ep_3}\dl(E_1-E_3)<g_{\ep_1},P_{E_1}g_{\ep_1}>
 N^{\iota}_{\ep_4,\ep_2,\la}(E_4,E_2,t')& \nonumber\\
 \vphantom{\ex}
 +\la^2\dl_{\ep_2,\ep_4}\dl(E_2-E_4)<g_{\ep_2},P_{E_2}L^2g_{\ep_2}>
 N_{\ep_3,\ep_1,\la}(E_3,E_1,t')\Bigr) \label{cr} \\
 \fl [B_{\ep_1,\ep_2,\la}(E_1,E_2,t),N_{\ep_3,\ep_4,\la}(E_3,E_4,t')]\nonumber\\
 =\dl_{\ep_1,\ep_3}\ex\dl(E_1-E_3)\nonumber\\
 \times <g_{\ep_1},P_{E_1}g_{\ep_1}>
 B_{\ep_4,\ep_2,\la}(E_4,E_2,t') \label{cr2} \\
 \fl [N_{\ep_1,\ep_2,\la}(E_1,E_2,t),N_{\ep_3,\ep_4,\la}(E_3,E_4,t')]\nonumber\\
 =\frac{\rme^{\rmi (t'-t)(E_3-E_1)/\la^2}}{\la^2}
 \Bigl(\dl_{\ep_2,\ep_3}\dl(E_2-E_3)<g_{\ep_2},P_{E_2}g_{\ep_2}>
 N_{\ep_1,\ep_4,\la}(E_1,E_4,t') \nonumber \\
 \vphantom{\ex}
 -\dl_{\ep_1,\ep_4}\dl(E_1-E_4)<g_{\ep_1},P_{E_1}g_{\ep_1}>{}
 N_{\ep_3,\ep_2,\la}(E_3,E_2,t)\Bigr).\label{cr3}
\end{eqnarray}

Notice that in the sense of distributions one has the limit
\begin{equation}\label{dllimit}
 \lim\limits_{\la\to 0}\ex = 2\pi\dl(t'-t)\dl(E_1-E_2)
\end{equation}
and, in the sense of distributions over the standard simplex
(cf.~\cite{ALV}) one has the limit
\begin{equation}\label{dlplimit}
 \lim\limits_{\la\to 0}\ex =\dl_+(t'-t)\frac{1}{\rmi (E_1-E_2-\rmi 0)}.
\end{equation}

The following theorem describes the algebra of commutation
relations for the master field in the LDL.
\begin{theorem}\label{mastfi}
The limits
\[
 X_{\ep_1,\ep_2}(E_1,E_2,t):=
 \lim\limits_{\la\to 0}X_{\ep_1,\ep_2,\la}(E_1,E_2,t)\qquad (X=B,N)
\]
exist in the sense of convergence of correlators
and satisfy the (causal) commutation relations
\begin{eqnarray}
 \fl [B_{\ep_1,\ep_2}(E_1,E_2,t),B^+_{\ep_3,\ep_4}(E_3,E_4,t')] \nonumber\\
 \vphantom{\frac{P}{E_1}}
 = \dl_{\ep_1,\ep_3}\dl_{\ep_2,\ep_4}\dl_+(t'-t)\dl(E_1-E_3)\dl(E_2-E_4)
 \nonumber\\
 \times\frac{<g_{\ep_1},P_{E_1}g_{\ep_1}>}{\rmi (E_1-E_2-\rmi 0)}
 <g_{\ep_2},P_{E_2}L^2g_{\ep_2}>\label{crmf1} \\
 \fl\vphantom{\frac{P}{E_1}}
 [B_{\ep_1,\ep_2}(E_1,E_2,t),N_{\ep_3,\ep_4}(E_3,E_4,t')]  \nonumber\\
 = \dl_{\ep_1,\ep_3}\dl_+(t'-t)\dl(E_1-E_3)\frac{<g_{\ep_1},P_{E_1}g_{\ep_1}>}
 {\rmi (E_1-E_2-\rmi 0)}B_{\ep_4,\ep_2}(E_4,E_2,t') \label{crmf2} \\
 \fl [N_{\ep_1,\ep_2}(E_1,E_2,t),N_{\ep_3,\ep_4}(E_3,E_4,t')]=
 \dl_+(t'-t)\frac{1}{\rmi (E_3-E_1-\rmi 0)} \nonumber \\
 \times\Bigl(\dl_{\ep_2,\ep_3}\dl(E_2-E_3)<g_{\ep_2},P_{E_2}g_{\ep_2}>
 N_{\ep_1,\ep_4}(E_1,E_4,t') \nonumber \\
 - \dl_{\ep_1,\ep_4}\dl(E_1-E_4)<g_{\ep_1},P_{E_1}g_{\ep_1}>{}
 N_{\ep_3,\ep_2}(E_3,E_2,t)\Bigr) \label{crmf3}
\end{eqnarray}
The commutation relations of the master field are obtained
by (\ref{crmf1}),(\ref{crmf2}),(\ref{crmf3}) replacing the factor
$ \dl_+(t'-t) $ by $ \dl(t'-t) $ and
$ (\rmi (E_1-E_2-\rmi 0))^{-1} $ by $ 2\pi\dl(E_1-E_2) $.
\end{theorem}
{\bf Proof.} The proof of the theorem follows by induction from
the commutation relations~(\ref{cr})-(\ref{cr3}) using the
limits~(\ref{dllimit}) and~(\ref{dlplimit}) and standard methods
of the stochastic limit.

%%%%%%%%%%%%%%%%%%%%%%%%%%%%%%%%%%%%%%%%%%%%%%%%%%%%%%%%%%%%%%%%%%%%%%%%%%%%%%%

\section{The master space and the associated white noise}

Let $ {\cal H}_{0,1} $ denote the closed subspace of
$ {\cal H}_1=L^2({\mathbb R}^d) $ spanned by the vectors
\[
 S_tg_\ep,\quad \ep\in\{0,1\},\quad t\in{\mathbb R}.
\]
Let $ K $ be a non zero subspace of $ {\cal H}_{0,1} $
such that $ g_\ep\in K $ ($ \ep=0,1 $) and
\[
 \int\limits_{-\infty}^\infty|<f,S_tg>|\rmd t<\infty\qquad\forall f,g\in K.
\]
This assumption implies that the sesquilinear form $ (\cdot|\cdot):K\times K
\longrightarrow C $ defined by
\[
 (f|g)=\int\limits_{-\infty}^\infty<f,S_tg>\rmd t, \qquad f,g\in K
\]
is well defined. Moreover it defines
a pre-scalar product on $ K $. We denote $ \{K,(\cdot|\cdot)\} $ or
simply $ K $, the completion of the quotient of $ K $ by the zero
$ (\cdot|\cdot) $-norm elements.

Define then Hilbert space $ K_{0,1}$ which will be denoted also as
$K\om_\beta K $ as the completion of the algebraic tensor product
$ K\om K^* $ with respect to the scalar product
\begin{eqnarray*}\fl
 (f_0\om_\beta f_1|f'_0\om_\beta f'_1)&:=\int\limits_{-\infty}^\infty
 <f_0,S_tf'_0>\overline{<f_1,S_tL^2f'_1>}\rmd t\\
 & = \int\limits_{-\infty}^\infty <f_0,S_tf'_0><f'_1,S_{-t}L^2f_1>\rmd t.
\end{eqnarray*}

Bounded operators acts naturally on $ K_{0,1} $ by
\[
 (A\om_\beta B)(f_0\om_\beta f_1)=Af_0\om_\beta Bf_1
 \qquad\forall A,B\in {\bf B}(K).
\]

The limit reservoir (or master) space is the space
\[
 {\cal F}(L^2({\mathbb R})\om K_{0,1}).
\]

The (non-causal) commutation relations
(\ref{crmf1}),...,(\ref{crmf3}) mean that operators
$ B_{\ep_1,\ep_2}(E_1,E_2,t) $
are the white noise operators $ b_t(\cdot) $ in
$ {\cal F}(L^2({\mathbb R})\om K_{0,1}) $:
\[
 B_{\ep_1,\ep_2}(E_1,E_2,t)=:b_t(P_{E_1}g_{\ep_1}\om_\beta P_{E_2}g_{\ep_2}).
\]

The number operator is
\begin{eqnarray}\fl
 N_{\ep_1,\ep_2}(E_1,E_2,t)=\sum\limits_{\ep'=0,1}n_{\ep'}(E_1)
 b^+_t(g_{\ep_1}\om_\beta P_{E_1}g_{\ep'})
 b_t(P_{E_2}g_{\ep_2}\om_\beta P_{E_1}g_{\ep'})\nonumber\\
 =\sum\limits_{\ep'=0,1}\int\rmd En_{\ep'}(E_1)
 B^+_{\ep_1,\ep'}(E,E_1,t)B_{\ep_2,\ep'}(E_2,E_1,t)\label{noperator}
\end{eqnarray}
where we denoted
\[
 n_\ep(E):=\frac{1}{<g_\ep,P_EL^2g_\ep>}.
\]

By identifying the element of the algebraic tensor product
$ f\om g\in K\om K^* $ with the operator
\[
 |f><g|:\xi\in{\cal H}_1\to<g,\xi>f\in{\cal H}_1
\]
so that
\[
 \bigl(|f><g|\bigr)^*=|g><f|
\]
and introducing the scalar product of such operators $ X,Y $
(notice that $ L^2=e^{-\beta H} $) by
\[
 <Y,X>:=Tr\int\rme^{-\beta H}Y^*S_tXS^*_t\rmd t=
        2\pi Tr\int\rmd E\rme^{-\beta H}Y^*P_EXP_E
\]
one can rewrite white noise $ b_t(g\om_\beta f) $ as $ b_t(|g><f|) $
with commutation relations defined by
\[
 [b_t(Y),b_{t'}^+(X)]=\dl(t'-t)<Y,X>.
\]

Let us introduce for simplicity
\begin{eqnarray*}
 B_{\ep_1,\ep_2}(E,t):=\int\rmd E'B_{\ep_1,\ep_2}(E',E,t)\\
 N_{\ep_1,\ep_2}(E,t):=\int\rmd E'N_{\ep_1,\ep_2}(E,E',t)
\end{eqnarray*}
with (causal) commutation relations:
\begin{eqnarray*}\fl
 [B_{\ep_1,\ep_2}(E,t),B^+_{\ep_3,\ep_4}(E',t')]=
 \dl_{\ep_1,\ep_3}\dl_{\ep_2,\ep_4}\dl_+(t'-t)\dl(E-E')\ga_{\ep_1}(E)
 <g_{\ep_2},P_EL^2g_{\ep_2}>\\
 \fl [B_{\ep_1,\ep_2}(E,t),N_{\ep_3,\ep_4}(E',t')]=
 \dl_{\ep_1,\ep_3}\dl_+(t'-t)\frac{<g_{\ep_1},P_{E'}g_{\ep_1}>}
 {\rmi (E'-E-\rmi 0)}B_{\ep_4,\ep_2}(E,t').
\end{eqnarray*}

In these notations the limiting Hamiltonian
acts on $ {\cal H}_S\om{\cal F}(L^2({\mathbb R})\om K_{0,1}) $ as
\[
 H(t)=\int\rmd E\sum\limits_{\ep=0,1}D_\ep\om\bigl(
 N_{\ep,1-\ep}(E,t)+B_{1-\ep,\ep}(E,t)+B_{\ep,1-\ep}^+(E,t)\bigr).
\]

%%%%%%%%%%%%%%%%%%%%%%%%%%%%%%%%%%%%%%%%%%%%%%%%%%%%%%%%%%%%%%%%%%%%%%%%%%%%%%%

\section{Emergence of the drift term and annihilation process}

The results of the preceding section allow us to write the
equation for the evolution operator in the stochastic limit
\begin{equation}\label{equ}\fl
 \partial_tU_t=-\rmi H(t)U_t=-\rmi\sum\limits_{\ep=0,1}D_\ep\om(N_{\ep,1-\ep}(t)+
 B_{1-\ep,\ep}(t)+B_{\ep,1-\ep}^+(t))U_t
\end{equation}
In order to bring it to the normally ordered form one needs to compute
the commutator
\[
 -\rmi D_\ep[B_{1-\ep,\ep}(t),U_t]=-\rmi D_\ep\int\rmd E[B_{1-\ep,\ep}(E,t),U_t].
\]

Notice that $ D_\ep D_{1-\ep} $ is a positive self-adjoint operator.
Therefore one can assume that for each $ E\in {\mathbb R} $,
the inverse operator
\[
 T_\ep(E):=\bigl(1+(\ga_\ep\ga_{1-\ep})(E)D_\ep D_{1-\ep}\bigr)^{-1}
\]
always exists. Notice also that, since $ D_\ep D_{1-\ep} $
commutes with $ T_\ep(E) $, one has
\[
 1-D_\ep D_{1-\ep}(\ga_\ep\ga_{1-\ep})(E)T_\ep(E)=T_\ep(E).
\]
Therefore
\begin{equation}\label{TE}
 \rmi D_\ep(1-D_{1-\ep}(\ga_\ep\ga_{1-\ep})(E)T_\ep(E)D_\ep)=\rmi T_\ep(E)D_\ep
\end{equation}
\begin{theorem}\label{normordu}
For the model described above one has
\begin{eqnarray}\fl
 -\rmi D_\ep[B_{1-\ep,\ep}(t),U_t]=-\int\rmd E\ga_{1-\ep}(E)D_\ep D_{1-\ep}T_\ep(E)\nonumber\\
 \times\Bigl(<g_\ep,P_EL^2g_\ep>U_t-
 \rmi D_\ep\ga_\ep(E)U_tB_{1-\ep,\ep}(E,t)+U_tB_{\ep,\ep}(E,t)\Bigr)\label{commbu}
\end{eqnarray}
\end{theorem}
{\bf Proof.} Using the integral equation for the evolution operator
and the commutation relations~(\ref{crmf1}),(\ref{crmf2}), one gets
\begin{eqnarray}\fl
 -\rmi D_\ep[B_{1-\ep,\ep}(E,t),U_t]\nonumber\\
 \fl =-\sum\limits_{\ep'=0,1}D_\ep D_{\ep'}
 \int\rmd E'\int\limits_0^t\rmd t_1[B_{1-\ep,\ep}(E,t),
 N_{\ep',1-\ep'}(E',t_1)+B^+_{\ep',1-\ep'}(E',t_1)]U_{t_1}\nonumber\\
 =-D_{\ep}D_{1-\ep}\ga_{1-\ep}(E)\Bigl(
 <g_\ep,P_{E}L^2g_\ep>+B_{\ep,\ep}(E,t)\Bigr)U_t\label{commbu1}
\end{eqnarray}
Notice that the first equality in~(\ref{commbu1}) holds
because, due to the time consecutive principle
\[
 [B_{\ep,\ep'}(E,t),U_{t_1}]=0.
\]
Similarly one computes the commutator
\begin{eqnarray}\fl
 [B_{\ep,\ep}(E,t),U_t]=-\rmi\sum\limits_{\ep'}D_{\ep'}\int\rmd E'
 \int\limits_0^t\rmd t_1[B_{\ep,\ep}(E,t),N_{\ep',1-\ep'}(E',t_1)]U_{t_1}\nonumber\\
 = -\rmi D_\ep\ga_\ep(E)B_{1-\ep,\ep}(E,t)U_t\label{commbu1a}
\end{eqnarray}
After substitution of this commutator into (\ref{commbu1}) one gets
\begin{eqnarray*}\fl
 -\rmi D_\ep[B_{1-\ep,\ep}(E,t),U_t]=-D_{\ep}D_{1-\ep}\ga_{1-\ep}(E)
 \bigl(<g_\ep,P_EL^2g_\ep>U_t-\rmi D_\ep\ga_\ep(E)\\
 \times ([B_{1-\ep,\ep}(E,t),U_t]+U_tB_{1-\ep,\ep}(E,t))+U_tB_{\ep,\ep}(E,t)\bigr)
\end{eqnarray*}
Then for
\[
 f_\ep(E,t):=-\rmi D_\ep[B_{1-\ep,\ep}(E,t),U_t]
\]
one has
\begin{eqnarray*}
 \fl\bigl(1+(\ga_\ep\ga_{1-\ep})(E)D_\ep D_{1-\ep}\bigr)f_\ep(E,t)
 =-\ga_{1-\ep}(E)D_\ep D_{1-\ep}\bigl(<g_\ep,P_EL^2g_\ep>U_t \\
 -\rmi D_\ep\ga_\ep(E)U_tB_{1-\ep,\ep}(E,t)+U_tB_{\ep,\ep}(E,t)\bigr).
\end{eqnarray*}

Since the inverse operator
$ \bigl(1+(\ga_\ep\ga_{1-\ep})(E)D_\ep D_{1-\ep}\bigr)^{-1} $ exists
we can solve the equation above for $ f_\ep(E,t) $. Using this
solution we find
\begin{eqnarray*}\fl
 -\rmi D_\ep[B_{1-\ep,\ep}(t),U_t]=\int\rmd Ef_\ep(E,t)=
 -\int\rmd E\ga_{1-\ep}(E)D_\ep D_{1-\ep}T_\ep(E)\\
 \times\Bigl(<g_\ep,P_EL^2g_\ep>U_t
 -\rmi D_\ep\ga_\ep(E)U_tB_{1-\ep,\ep}(E,t)+U_tB_{\ep,\ep}(E,t)\Bigr).
\end{eqnarray*}

%%%%%%%%%%%%%%%%%%%%%%%%%%%%%%%%%%%%%%%%%%%%%%%%%%%%%%%%%%%%%%%%%%%%%%%%%%%%%%

\section{Emergence of the number and creation processes}

In order to bring equation (\ref{equ}) to the normally ordered form
one needs also to move the annihilation operators in $ N_{\ep,1-\ep}(t) $ to the
right of the evolution operator. Using~(\ref{noperator}) this leads to
\begin{eqnarray}\fl
 -\rmi D_\ep N_{\ep,1-\ep}(t)U_t=-\rmi D_\ep\int\rmd E\sum\limits_{\ep'=0,1}
 n_{\ep'}(E)B^+_{\ep,\ep'}(E,t)B_{1-\ep,\ep'}(E,t)U_t\nonumber\\
 \fl = -\rmi D_\ep\int\rmd E\sum\limits_{\ep'=0,1}n_{\ep'}(E)B^+_{\ep,\ep'}(E,t)
 \Bigl([B_{1-\ep,\ep'}(E,t),U_t]+U_tB_{1-\ep,\ep'}(E,t)\Bigr)\label{normordu1}
\end{eqnarray}
and the commutator is evaluated using~(\ref{commbu1}) and (\ref{commbu1a}).
\begin{theorem}\label{normordu2} For the model described above one has
\begin{eqnarray*}\fl
 -\rmi D_\ep N_{\ep,1-\ep}(t)U_t=-\int\rmd E\Bigl(\ga_{1-\ep}(E)D_\ep
 D_{1-\ep}T_\ep(E)\\
 \times\Bigl(-\rmi D_\ep\ga_\ep(E)B^+_{\ep,1-\ep}(E,t)+B^+_{\ep,\ep}(E,t)\Bigr)U_t\\
 +\sum\limits_{\ep'=0,1}n_{\ep'}(E)
 \Bigl(\rmi T_\ep(E)D_\ep B^+_{\ep,\ep'}(E,t)U_tB_{1-\ep,\ep'}(E,t)\\
 +\ga_{1-\ep}(E)D_\ep D_{1-\ep}T_\ep(E)B^+_{\ep,\ep'}(E,t)U_tB_{\ep,\ep'}(E,t)\Bigr)\Bigr)
\end{eqnarray*}
\end{theorem}
{\bf Proof.} From~(\ref{commbu1}) and (\ref{commbu1a}) it follows that
\begin{eqnarray*}\fl
 -\rmi D_\ep[B_{1-\ep,\ep'}(E,t),U_t]=-\ga_{1-\ep}(E)D_\ep D_{1-\ep}T_\ep(E)
 \Bigl(n^{-1}_{\ep'}(E)(\dl_{\ep,\ep'}-\dl_{1-\ep,\ep'}\rmi D_\ep\ga_\ep)U_t \\
 -\rmi D_\ep\ga_\ep(E)U_tB_{1-\ep,\ep'}(E,t)+U_tB_{\ep,\ep'}(E,t)\Bigr).
\end{eqnarray*}
After substitution of these commutators in~(\ref{normordu1})
and using~(\ref{TE}) one finishes the proof of the theorem.

%%%%%%%%%%%%%%%%%%%%%%%%%%%%%%%%%%%%%%%%%%%%%%%%%%%%%%%%%%%%%%%%%%%%%%%%%%%%%%%

\section{The normally ordered equation}

Theorems~(\ref{normordu}) and~(\ref{normordu2})
allow us to obtain immediately the normally ordered equation for the
evolution operator in the LDL. This procedure of deduction of
quantum stochastic differential equation is being called a stochastic golden
rule. The normally ordered equation has the form
\begin{eqnarray}\fl
 \partial_tU_t=-\sum\limits_{\ep=0,1}\int\rmd E\Bigl\{
 \rmi T_\ep(E)D_\ep\sum\limits_{\ep'=0,1}
 n_{\ep'}(E)B^+_{\ep,\ep'}(E,t)U_tB_{1-\ep,\ep'}(E,t)\nonumber\\
 +\vphantom{\sum\limits_0\int}
 \ga_{1-\ep}(E)D_\ep D_{1-\ep}T_\ep(E)\sum\limits_{\ep'=0,1}
 n_{\ep'}(E)B^+_{\ep,\ep'}(E,t)U_tB_{\ep,\ep'}(E,t)\nonumber\\
 +\vphantom{\sum\limits_0\int}
 \rmi T_\ep(E)D_\ep B^+_{\ep,1-\ep}(E,t)U_t+
 \ga_{1-\ep}(E)D_\ep D_{1-\ep}T_\ep(E)B^+_{\ep,\ep}(E,t)U_t\nonumber\\
 +\vphantom{\sum\limits_0\int}
 \rmi T_\ep(E)D_\ep U_tB_{1-\ep,\ep}(E,t)+
 \ga_{1-\ep}(E)D_\ep D_{1-\ep}T_\ep(E)U_tB_{\ep,\ep}(E,t)\nonumber\\
 +\vphantom{\sum\limits_0\int}
 \ga_{1-\ep}(E)D_\ep D_{1-\ep}T_\ep(E)<g_\ep,P_EL^2g_\ep>U_t\Bigr\}\label{normordeq}
\end{eqnarray}
We will represent equation~(\ref{normordeq}) in the form of
quantum stochastic differential equation~\cite{HP}.
\begin{theorem}
Equation~(\ref{normordeq}) is equivalent to the
quantum stochastic differential equation
\begin{eqnarray}\fl
 \rmd U_t=-\sum\limits_{\ep=0,1}\int\rmd E\Bigl\{
 \rmi T_\ep(E)D_\ep\rmd N_t(2\pi |g_\ep><g_{1-\ep}|P_E\om_\beta P_E)\nonumber\\
 +\vphantom{\sum\limits_0\int}
 \ga_{1-\ep}(E)D_\ep D_{1-\ep}T_\ep(E)
 \rmd N_t(2\pi |g_\ep><g_\ep|P_E\om_\beta P_E)\nonumber\\
 +\vphantom{\sum\limits_0\int}
 \rmi T_\ep(E)D_\ep\rmd B^+_t(g_\ep\om_\beta P_Eg_{1-\ep})+
 \ga_{1-\ep}(E)D_\ep D_{1-\ep}T_\ep(E)\rmd B^+_t(g_\ep\om_\beta P_Eg_\ep)\nonumber\\
 +\vphantom{\sum\limits_0\int}
 \rmi T_\ep(E)D_\ep\rmd B_t(g_{1-\ep}\om_\beta P_Eg_\ep)+
 \ga_{1-\ep}(E)D_\ep D_{1-\ep}T_\ep(E)\rmd B_t(g_\ep\om_\beta P_Eg_\ep)\nonumber\\
 +\vphantom{\sum\limits_0\int}
 \ga_{1-\ep}(E)D_\ep D_{1-\ep}T_\ep(E)<g_\ep,P_EL^2g_\ep>
 \rmd t\Bigr\}U_t\label{normordeqa}
\end{eqnarray}
\end{theorem}
{\bf Proof.} Let us consider the following term in~(\ref{normordeq})
\begin{eqnarray}\fl
 \sum\limits_{\ep'=0,1}n_{\ep'}(E)B^+_{\ep,\ep'}(E,t)U_tB_{1-\ep,\ep'}(E,t)=
 n_{1-\ep}(E)b^+_t(g_\ep\om_\beta P_Eg_{1-\ep})U_tb_t(g_{1-\ep}\om_\beta
 P_Eg_{1-\ep})\nonumber\\
 +n_\ep(E)b^+_t(g_\ep\om_\beta P_Eg_\ep)U_tb_t(g_{1-\ep}\om_\beta
 P_Eg_\ep)\label{1N}
\end{eqnarray}
The matrix element of this expression on the exponential vectors $ \psi(f) $,
$ \psi(f') $ is (we use Dirac's notations also for bra- and ket-vectors
from $ K_{0,1} $)
\begin{eqnarray}\fl
 \bigl(n_{1-\ep}(E)<f|g_\ep\om_\beta P_Eg_{1-\ep}><g_{1-\ep}\om_\beta
 P_Eg_{1-\ep}|f'>\nonumber\\
 +n_\ep(E)<f|g_\ep\om_\beta P_Eg_\ep><g_{1-\ep}\om_\beta P_Eg_\ep|f'>\bigr)
 <\psi(f)|U_t|\psi(f')>\nonumber\\
 =<f|T_{\ep,1-\ep}(E)|f'><\psi(f)|U_t|\psi(f')>\label{1Na}
\end{eqnarray}
which is the time derivative of the matrix element
$ <\psi(f)|\rmd N_t(T_{\ep,1-\ep}(E))|\psi(f')> $
where $ N_t(T_{\ep,1-\ep}(E)) $ is the number process with intensity
\begin{eqnarray*}\fl
 T_{\ep,1-\ep}(E)=n_{1-\ep}(E)|g_\ep\om_\beta P_Eg_{1-\ep}><g_{1-\ep}\om_\beta
 P_Eg_{1-\ep}|\\
 +n_\ep(E)|g_\ep\om_\beta P_Eg_\ep><g_{1-\ep}\om_\beta P_Eg_\ep|
\end{eqnarray*}
Let us now prove that
\begin{equation}
 T_{\ep,1-\ep}(E)=2\pi |g_\ep><g_{1-\ep}|P_E\om_\beta P_E\label{1Nb}
\end{equation}
To this goal let us consider the action of the
$ T_{\ep,1-\ep}(E) $ on vectors of the form
\[
 |f>=|P_{E_1}g_{\ep_1}\om_\beta P_{E_2}g_{\ep_2}>.
\]
One has
\begin{eqnarray}\fl
 T_{\ep,1-\ep}(E)|f>\nonumber\\
 =2\pi\dl_{\ep,1-\ep_1}\dl(E_1-E)\dl(E_2-E)<g_{1-\ep},P_Eg_{1-\ep}>\nonumber\\
 (\dl_{\ep,1-\ep_2}|g_\ep\om_\beta P_Eg_{1-\ep}>+
 \dl_{\ep,\ep_2}|g_\ep\om_\beta P_Eg_\ep>)\nonumber\\
 =2\pi\dl_{\ep,1-\ep_1}<g_{1-\ep},P_Eg_{1-\ep}>\dl(E_1-E)\dl(E_2-E)\nonumber\\
 (\dl_{\ep,1-\ep_2}|g_\ep\om_\beta P_Eg_{\ep_2}>+
 \dl_{\ep,\ep_2}|g_\ep\om_\beta P_Eg_{\ep_2}>)\nonumber\\
 =2\pi\dl_{\ep,1-\ep_1}<g_{1-\ep},P_Eg_{1-\ep}>\dl(E_1-E)\dl(E_2-E)
 |g_\ep\om_\beta P_Eg_{\ep_2}>\nonumber\\
 =2\pi |g_\ep><g_{1-\ep}|P_E\om_\beta P_E|f>\label{1Nc}
\end{eqnarray}
Therefore (\ref{1Nb}) holds and
the term~(\ref{1N}) corresponds to the number process
\[
 \rmd N_t(2\pi |g_\ep><g_{1-\ep}|P_E\om_\beta P_E)
\]

Computing the same matrix element for the term
\begin{eqnarray}\fl
 \sum\limits_{\ep'=0,1}n_{\ep'}(E)B^+_{\ep,\ep'}(E,t)U_tB_{\ep,\ep'}(E,t)=
 n_{1-\ep}(E)b^+_t(g_\ep\om_\beta P_Eg_{1-\ep})U_tb_t(g_\ep\om_\beta
 P_Eg_{1-\ep})\nonumber\\
 +n_\ep(E)b^+_t(g_\ep\om_\beta P_Eg_\ep)U_tb_t(g_\ep\om_\beta P_Eg_\ep)\label{2N}
\end{eqnarray}
one finds an expression like~(\ref{1Na}) with $ T_{\ep,1-\ep} $
replaced by the operator
\[\fl
 T_{\ep,\ep}(E)=n_{1-\ep}(E)|g_\ep\om_\beta P_Eg_{1-\ep}><g_\ep\om_\beta
 P_Eg_{1-\ep}|+n_\ep(E)|g_\ep\om_\beta P_Eg_\ep><g_\ep\om_\beta P_Eg_\ep|
\]
and a calculation similar to the one done in~(\ref{1Nc}) leads to the
conclusion that
\[
 T_{\ep,\ep}(E)=2\pi |g_\ep><g_\ep|P_E\om_\beta P_E.
\]
Therefore the term~(\ref{2N}) corresponds to the number process
\[
 \rmd N_t(2\pi |g_\ep><g_\ep|P_E\om_\beta P_E).
\]
This finishes the proof of the theorem.

Let us introduce the notations:
\begin{eqnarray*}
 R_{\ep,\ep}(E)=-\ga_{1-\ep}(E)D_\ep D_{1-\ep}T_\ep(E)\\
 R_{\ep,1-\ep}(E)=-\rmi T_\ep(E)D_\ep .
\end{eqnarray*}
In these notations the quantum stochastic differential equation for the
evolution operator can be rewritten as
\begin{eqnarray*}\fl
 \rmd U_t=\sum\limits_{\ep=0,1}\int\rmd E\Bigl\{
 \vphantom{\sum\limits_0\int}
 R_{\ep,1-\ep}(E)\rmd N_t(2\pi|g_\ep><g_{1-\ep}|P_E\om_\beta P_E)\\
 +\vphantom{\sum\limits_0\int}
 R_{\ep,\ep}(E)\rmd N_t(2\pi|g_\ep><g_\ep|P_E\om_\beta P_E)\\
 +\vphantom{\sum\limits_0\int}
 R_{\ep,1-\ep}(E)\rmd B^+_t(g_\ep\om_\beta P_Eg_{1-\ep})+
 R_{\ep,\ep}(E)\rmd B^+_t(g_\ep\om_\beta P_Eg_\ep)\\
 +\vphantom{\sum\limits_0\int}
 R_{\ep,1-\ep}(E)\rmd B_t(g_{1-\ep}\om_\beta P_Eg_\ep)+
 R_{\ep,\ep}(E)\rmd B_t(g_\ep\om_\beta P_Eg_\ep)\\
 +\vphantom{\sum\limits_0\int}
 R_{\ep,\ep}(E)<g_\ep,P_EL^2g_\ep>\rmd t\Bigr\}U_t.
\end{eqnarray*}

Notice that the quantum stochastic differential
equation~(\ref{normordeqa}) can be written also in the
Frigerio-Maassen form~\cite{FrMa}. In order to prove this recall
that for any pair of Hilbert spaces $ {\cal X}_0,{\cal X}_1 $ if $ N,A $
denote the number and annihilation processes on the Fock space
$ {\cal F}({\cal X}_1) $ then for $ X_0\in B({\cal X}_0) $,
$ X_1\in B({\cal X}_1) $, $ x\in{\cal X}_1 $, Frigerio and Maassen~\cite{FrMa}
introduced the notation:
\begin{eqnarray*}
 N(X_0\om X_1):=X_0\om N(X_1)\\
 A(X_0\om X_1x):=X_0\om A(X_1x)\\
 <x,X_0\om X_1x>:=X_0\om 1<x,X_1x>
\end{eqnarray*}
Let us also introduce an operator $ T_3(E) $ acting on the triple
$ {\cal H}_S\om K\om_\beta K $ (this is the reason for introducing index $3$)
as
\[
 T_3(E):=2\pi\sum\limits_{\ep,\ep'=0,1}R_{\ep,\ep'}(E)
 \om|g_\ep><g_{\ep'}|P_E\om_\beta P_E
\]
and the vector $ \xi(E)\in K\om_\beta K $
\[
 \xi(E):=\frac{1}{2\pi}\sum\limits_{\ep=0,1}
 \frac{1}{<g_\ep,P_Eg_\ep>}g_\ep\om_\beta g_\ep .
\]

In these notations equation~(\ref{normordeqa}) can be written as
\begin{eqnarray*}\fl
 \rmd U_t=\int\rmd E\Bigl(\rmd N_t\bigl(T_3(E)\bigr)+\rmd B^+_t\bigl(T_3(E)\xi(E)\bigr)\\
 +\rmd B_t\bigl(T^*_3(E)\xi(E)\bigr)+<\xi(E),T_3(E)\xi(E)>\rmd t\Bigr)U_t.
\end{eqnarray*}

%%%%%%%%%%%%%%%%%%%%%%%%%%%%%%%%%%%%%%%%%%%%%%%%%%%%%%%%%%%%%%%%%%%%%%%%%%%%%%%

\section{Connection with scattering theory}

Here we consider relation between the evolution operator and scattering
theory.
Because of the number conservation,
the closed subspace of $ {\cal H}_S\om {\cal F} $ generated by vectors of the
form $ u\om A^+(f)\Phi $\enskip ($ u\in {\cal H}_{\rm S} $,
$ f\in {\cal H}_1=L^2({\mathbb R}^d) $) which is naturally isomorphic to
$ {\cal H}_{\rm S}\om {\cal H}_1 $, is globally invariant under the time evolution
operator $ \exp[\rmi (H_{\rm S}\om 1+1\om H_{\rm R}+V)t] $. Therefore the restriction of the time
evolution operator to this subspace corresponds to an evolution operator on
$ {\cal H}_{\rm S}\om {\cal H}_1 $ given by
\[
 \exp[\rmi(H_{\rm S}\om 1+1\om H_1+V_1)t]
\]
where
\begin{equation}\label{defV}
 V_1=\sum\limits_{\ep=0,1}D_\ep\om|g_\ep><g_{1-\ep}|
\end{equation}

The 1-particle M\o ller wave operators are defined as
\[\fl
 \Omega_{\pm}=s-\lim\limits_{t\to\pm\infty}\exp[\rmi(H_{\rm S}\om 1+1\om H+V_1)t]
 \exp[-\rmi(H_{\rm S}\om 1+1\om H)t]
\]
and the 1-particle $ T $-operator is defined as
\begin{equation}\label{defT}
 T=V_1\Omega_+
\end{equation}
From (\ref{defV}) it follows that
\[
 \Omega_{\pm}=s-\lim\limits_{t\to\pm\infty}U^{(1)}_t
\]
where $ U_t^{(1)} $ is the solution of
\[
 \partial_tU_t^{(1)}=-\rmi\left(\sum\limits_{\ep=0,1}
 D_\ep\om|S_tg_\ep><S_tg_{1-\ep}|\right)U_t^{(1)}\qquad U_0^{(1)}=1.
\]

In order to make a connection between the stochastic process $ U_t $
and scattering theory notice that the operator $ T_3(E) $ can be written as
\[
 T_3(E)=2\pi T(E)\om_\beta P_E
\]
where operator $ T(E) $ acts on $ {\cal H}_{\rm S}\om K $ as
\[
 T(E)=\sum\limits_{\ep,\ep'=0,1}R_{\ep,\ep'}(E)\om |g_\ep><g_{\ep'}|P_E.
\]
In~\cite{AcLu} it was proved that $ T $-operator defined by (\ref{defT})
connected with $ T(E) $ by the following formula
\[
 T=\int\rmd ET(E).
\]

%%%%%%%%%%%%%%%%%%%%%%%%%%%%%%%%%%%%%%%%%%%%%%%%%%%%%%%%%%%%%%%%%%%%%%%%%

\section{Vacuum expectation value}

For the vacuum matrix element of the evolution
operator
from~(\ref{normordeq}) one immediately gets
\begin{equation}\label{expvalue}
<U(t)>_{vac}=\rme^{-\Gamma t}.
\end{equation}
The operator $ \Gamma $ acts in $ {\cal H}_{\rm S} $ as
\[
\Gamma =\sum\limits_{\ep=0,1}\int
\rmd E\ga_{1-\ep}(E)D_{\ep}D_{1-\ep}T_{\ep}(E)<g_\ep,P_EL^2g_\ep>.
\]
\begin{theorem}
The operator $ \Gamma $ has a
non-negative real part (i.e. this operator describes
the damping).
\end{theorem}
{\bf Proof.} From the definition of $T_\ep(E)$ we know that
\[
\gamma_{1-\ep}(E)D_\ep D_{1-\ep}T_\ep (E)={D_\ep
D_{1-\ep}\over{\gamma^{-1}_{1-\ep}(E)}+\gamma_\ep (E)
D_\ep D_{1-\ep}}
\]
But $\gamma_\ep$ and $\gamma_{1-\ep}$
(hence also $1/\gamma_{1-\ep}$) have positive real part
and $D_\ep D_{1-\ep}$ is positive self--adjoint.

Hence the above expression has a positive real part
because it is of the form:
\[
{H\over z_1+z_2H}\,={H(\hbox{Re }z_1+H\hbox{ Re
}z_2)\over|z_1+
z_2H|^2}\,-\rmi{H(\hbox{Im }z_1+H\hbox{ Im
}z_2)\over|z_1+z_2H|^2}
\]
where $H$ is positive self--adjoint and $z_1$, $z_2$
have a positive
real part. Since $\langle g_\ep,P_EL^2g_\ep\rangle\geq0$ the thesis follows.

%%%%%%%%%%%%%%%%%%%%%%%%%%%%%%%%%%%%%%%%%%%%%%%%%%%%%%%%%%%%%%%%%%%%%%%

\ack{
One of the authors (A.P.) is grateful to the Centro Vito Volterra for kind
hospitality, and to Y.G.Lu for useful discussions. This work is partially
supported by the RFFI 02-01-01084 and INTAS 99-00545 for L.A. and I.V. and by the
Vito Volterra Fellowship and INTAS 01/1-200 for A.P.}


\section*{References}
\begin{thebibliography}{99}

 \bibitem{AFL} Accardi L, Frigerio A and Lu Y G 1987
         {\it Quantum Probability and Applications. IV. Lect. Notes Math.}
         {\bf 1396} 20--58

 \bibitem{AKV} Accardi L, Kozyrev S V and Volovich I V 1997
         {\it Phys. Rev. A} {\bf 56}(4) 2557--62\\
         (Accardi L, Kozyrev S V and Volovich I V 1997 {\it Preprint} quant-ph/9706021)

 \bibitem{ALV} Accardi L, Lu Y G and Volovich I V 2001
         {\it Quantum Theory and Its Stochastic Limit}, Springer

 \bibitem{lanford} Lanford O E 1976 {\it Ast\'erisque}
         {\bf 40} 117--37

 \bibitem{grad} Grad H 1958 {\it Principles of the kinetic theory
         of gases. In: Handbuch der physik} {\bf 12} Fl\"ugge S (ed.),
         Berlin, Heidelberg, New York: Springer

 \bibitem{king} King F 1975 {\it PhD Thesis} University of
          California, Berkeley

 \bibitem{spohn2} Spohn H 1980 {\it Rewievs in Modern Physics}
         {\bf 52} 569--615

 \bibitem{dumcke} D\"umcke R 1984 {\it Lect. Notes in Math.}
         {\bf 1136} 151--61

 \bibitem{AcLu} Accardi L and Lu Y G 1991
         {\it J. Phys. A: Math. Gen.} {\bf 24} 3483--512

 \bibitem{kumerrer} Kummerer B 1986 {\it Markov dilations and non-commutative Poisson
         processes} Preprint Tubingen

 \bibitem{bogol} Bogoliubov N N 1946 {\it Porblems of Dynamical
          Theory in Statistical Physics}, (Moscow: Gostechizdat)

 \bibitem{davies} Davies E B 1974 {\it Comm. Math. Phys.}
         {\bf 39} 91--110

 \bibitem{Pl} Palmer P F 1977 {\it PhD Thesis} Oxford University

 \bibitem{HP} Hudson R and Parthasarathy K R 1984 {\it Comm. Math. Phys.}
         {\bf 93} 301--23

 \bibitem{FrMa} Frigerio A and Maassen H 1989 {\it Prob. Th. Rel. Fields}
         {\bf 83} 489--508


\end{thebibliography}
\end{document}